\documentclass[%
 reprint,
superscriptaddress,
bibnotes,
amsmath,amssymb,
aps,
pra,
]{revtex4-2}
\usepackage{graphicx}
\usepackage[caption=false]{subfig}
\usepackage{dcolumn}
\usepackage{bm}
\usepackage[loose,nice]{units}
\usepackage{xcolor}

\begin{document}

\preprint{APS/123-QED}

\title{Experimental evidence of a strange nonchaotic attractor in a Nd:YVO$_4$ laser with saturable absorber}

\author{Juliana N. Bourdieu}
\email{jbourdieu@citedef.gob.ar}
\affiliation{Laboratorio de Láseres Sólidos, Centro de Investigación en Láseres y Aplicaciones (CEILAP), Instituto de Investigaciones Científicas y Técnicas para la Defensa (CITEDEF), Villa Martelli, Buenos Aires, Argentina.}
\author{Mónica B.Agüero}

\affiliation{Laboratorio de Láseres Sólidos, Centro de Investigación en Láseres y Aplicaciones (CEILAP), Instituto de Investigaciones Científicas y Técnicas para la Defensa (CITEDEF), Villa Martelli, Buenos Aires, Argentina.}
\affiliation{Consejo Nacional de Investigaciones Científicas y Técnicas (CONICET), Argentina.}
\author{Marcelo G. Kovalsky}
\affiliation{Laboratorio de Láseres Sólidos, Centro de Investigación en Láseres y Aplicaciones (CEILAP), Instituto de Investigaciones Científicas y Técnicas para la Defensa (CITEDEF), Villa Martelli, Buenos Aires, Argentina.}
\affiliation{Consejo Nacional de Investigaciones Científicas y Técnicas (CONICET), Argentina.}

\date{\today}

\begin{abstract}

The diode-pumped Nd: Vanadate laser is one of the most widely used lasers in the near-infrared range. Using passive Q-switching with a Cr:YAG saturable absorber (SA), it generates nanosecond pulses at tens of kilohertz. It is well known to display either regular pulsing or low-dimensional deterministic chaos, depending on the absorber’s position
within the cavity. However, to our knowledge, a strange nonchaotic attractor (SNA) has never
been reported in this system. Our experiments provide the first evidence that, for a specific absorber position, the laser dynamics correspond to a strange nonchaotic attractor.

SNAs exhibit complex dynamical behavior that is neither periodic nor chaotic, 
positioning them as a valuable resource for applications requiring high spectral diversity while preserving predictable dynamics.

By experimentally obtaining time series of two laser variables, we reconstruct the underlying attractor. 
For a certain position of the SA, we find that all Lyapunov exponents are negative and that the energy spectra exhibit a broad frequency range, indicating that the attractor is nonchaotic and lacks periodicity.
Using Higuchi's method, we estimate the fractal dimension ($d_f$) and find that 13$\%$ of the nonchaotic time series exhibit a well-defined $d_f$, confirming the presence of an SNA. 

\end{abstract}

\maketitle

\section{Introduction} 
The study of attractors in nonlinear dynamical systems has been a cornerstone of chaos theory and complex systems research. While chaotic attractors, characterized by exponential sensitivity to initial conditions and fractal geometry, have been extensively studied since their first description by Ruelle and Takens \cite{Ruelle_Takens}, another intriguing class of attractors has emerged: strange nonchaotic attractors (SNAs). They were originally introduced in the 1980s by Grebogi et. al \cite{sna}. These attractors exhibit an unique combination of traits, merging the geometric complexity of strange attractors with the nonchaotic nature of systems characterized by non-positive Lyapunov exponents. Since their discovery, SNAs have been identified in various systems. In particular, experimental observations of SNAs have been documented in a magnetoelastic ribbon system under quasiperiodic driving \cite{Ditto}, in electronic circuits
\cite{zhou,yang}, in a memristor-based oscillator \cite{premraj}, and in an electrochemical cell \cite{ruiz}. 
In complex optical systems, such as lasers, chaotic attractors have been extensively studied both theoretically \cite{Roy} and experimentally \cite{Mandel}, since Haken \cite{haken} showed in 1975 that a single-mode laser is isomorphic to the Lorenz equations.
On the other hand, one of the most popular and simple methods for generating pulses in the nanosecond range, widely used in scientific fields, as well as in defense and industry \cite{Wei}, is the passively Q-switched Nd:Vanadate laser, whose dynamics are well known \cite{Din_1, Din_2}.
The Q-switched Nd:Vandate (Nd:YVO$_4$) laser with a saturable absorber has become a highly popular pulsed laser source due to its combination of high gain, excellent beam
quality, and compact, robust design.
Nd:YVO$_4$ offers efficient diode pumping and a high stimulated emission cross-section, \cite{RP} enabling the generation of short nanosecond pulses at kilohertz repetition rates without the need for complex electronic drivers.
Passive Q-switching with a saturable absorber, such as Cr:YAG, further enhances system simplicity, reliability, and cost-effectiveness.
These characteristics make the device well suited to a wide range of applications, including precision laser marking and engraving, micromachining of electronic components, medical and dermatological procedures, LIDAR and range-finding systems, and as a pump source for frequency conversion and other nonlinear optical processes.
However, strange nonchaotic attractors (SNA) had not been reported until now in this kind of laser. In this article, we present experimental evidence of SNA in a solid-state Nd:Vanadate laser with a Cr:YAG crystal as a saturable absorber. By employing experimental time series of two laser's output variables, pulse amplitude and time between consecutive pulses, we are able to reconstruct an equivalent attractor. A control parameter allows transitioning between the different dynamic behaviors of the laser.

\section{Setup and measurements}
\begin{figure}
\includegraphics[width=0.48\textwidth]{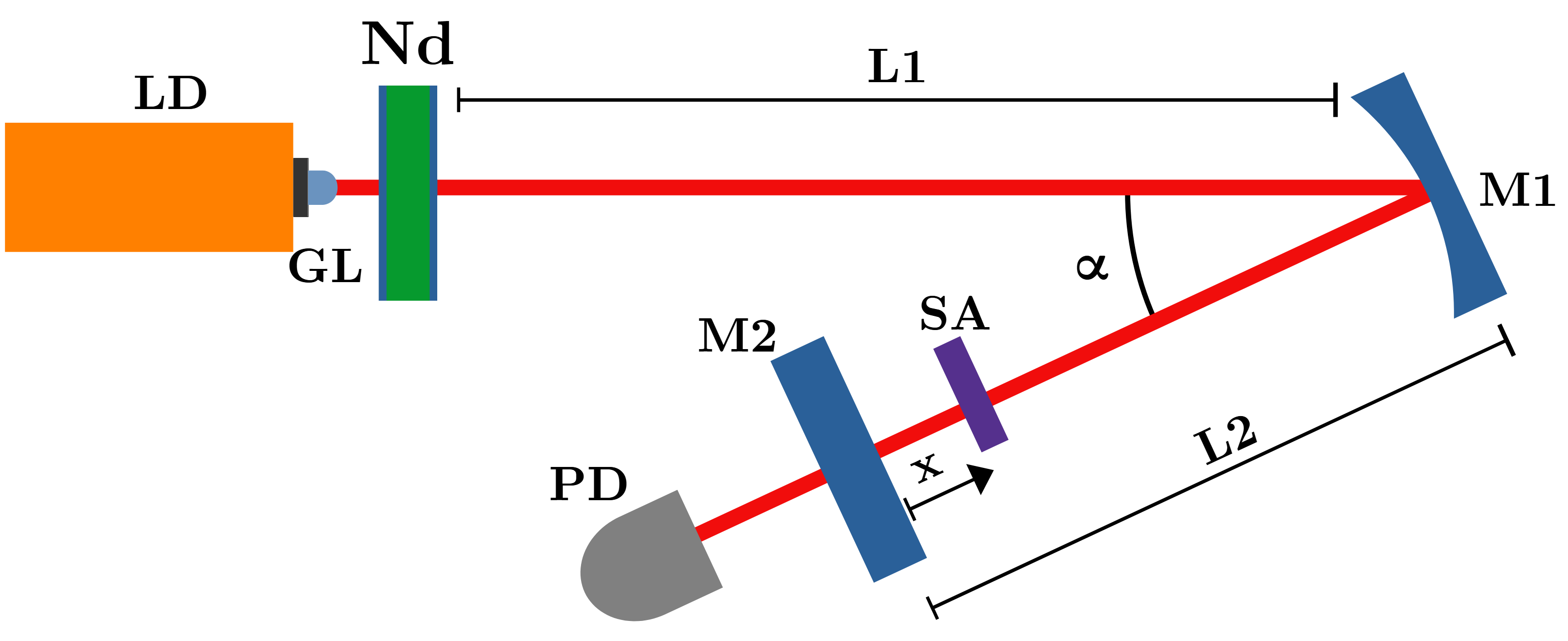}
\caption{\label{fig: setup} Experimental setup of the Q-switched Nd:VO$_4$ laser. LD: pump laser diode. GL: GRIN lens. Nd: Nd:VO$_4$ crystal slab. L1 and L2: lengths of the first and second cavity arms (\unit[12]{cm} and \unit[6]{cm}, respectively). $\alpha$: angle between the cavity arms ($\approx$ $19^\circ$). M1: HR concave mirror with a radius of curvature of \unit[10]{cm}. SA: Cr:YAG saturable absorber with 90$\%$ unbleached transmission. M2: plane output coupler with a reflectivity of 99$\%$. x: distance between M2 and SA. PD: fast photodiode. }
\end{figure}

Our Nd:VO$_4$ laser operates in a standard V-shaped cavity.
In Fig. \ref{fig: setup}, a schematic of our laser is shown.
The cavity consists of a high-reflectivity (HR) concave mirror placed at the folding point of the cavity with a radius of curvature of \unit[10]{cm}, and a plane output coupler with 99$\%$ reflectivity.
The operating wavelength of the laser is \unit[1064]{nm} and is linearly polarized.
A solid-state saturable absorber (SA), Cr:YAG, with 90$\%$ unbleached transmission, is placed in the second arm of the cavity, between the folding mirror and the output coupler.
The active medium is a \unit[3 × 3 × 1 ]{mm} Nd:VO$_4$ crystal, doped with 1$\%$, and longitudinally pumped by a \unit[2]{W} laser diode emitting at \unit[808]{nm}.
The laser is operated in conditions of normal use, well above threshold.

The control parameter is the energy density over the SA, which is adjusted by varying the SA position within the laser cavity.
Typical output of this laser is a pulse train with a repetition rate in the tens of kilohertz range and pulse duration of approximately \unit[150]{ns}. 
The dynamic regime of the laser output strongly depends on the SA position, going from stable Q-switch to chaos as the position is changed.

The laser output is recorded with a fast photodiode with a rise time of \unit[100]{ps}, connected to a digital storage oscilloscope (PicoScope 6403B: 350 MHz bandwidth, \unit[5]{GS/s} sampling rate, and \unit[1]{GS} memory).
The sampling frequency was chosen to guarantee an adequate temporal resolution of the laser dynamics.
By ensuring an average of 10 data points per pulse, and given that the mean pulse duration was 150 ns, a sampling rate of 70 MS/s is sufficient to reliably reconstruct the temporal profile of the signal.
We record the laser intensity over time while adjusting the SA position, resulting in 1,719 data sets, each containing between 6,000 and 50,800 pulses.
For each data set, we extracted two time series, one corresponding to the maximum intensity of each pulse and another with time between successive pulses, yielding  a total of 3,438 time series.

\section{Attractor reconstruction and classification}
The embedding theorem \cite{Takens, Mañé} states that an attractor's properties can be reconstructed from the time series of any system variable.
In our case, we apply these techniques to the amplitude and time difference series. 
For each variable, we construct vectors in a $d$-dimensional space using $s(n)$ and its time delays, $s(n+T)$, as follows:

\begin{multline}\label{eq:attractor}
    y(n) = \big[ s(n), s(n+T), s(n+2T), \ldots, \\
    s(n+(d-1)T) \big].
\end{multline}

By selecting the appropriate time delay and embedding dimension, $d_E$, the vector $y(n)$ reconstructs the phase space.
This means that all motion invariants observed in the reconstructed time-delay space match those in the original space.

We define the time delay of the phase space, $T$, as the first minimum of the mutual information \cite{Fraser}.
This criterion implies that the values of $s(n)$ and $s(n+T)$ are sufficiently independent to be used as coordinates in $y(n)$, while still maintaining some level of correlation.
To determine the embedding dimension, we use the false nearest neighbors (FNN) method \cite{Abarbanel}.

Using these tools, the underlying attractor of a time series can be reconstructed, and its dynamics classified by computing the Lyapunov exponents, $\left\{ \lambda_1,\lambda_2, \ldots, \lambda_{d_E}\right\}$, and the fractal dimension.

The signature of a chaotic attractor is the presence of at least, one positive Lyapunov exponent \cite{Abarbanel_book}. 
The maximum exponent, $\lambda_{max}$, quantifies the degree of sensitivity to initial conditions.
A negative sum of the Lyapunov exponents, $\sum_{i=1}^{d_E} \lambda_i$, indicates a dissipative system. 
Strange nonchaotic attractors require the computation of both the Lyapunov exponents and the fractal dimension. 
If both the sum of the Lyapunov exponents and $\lambda_{max}$ are negative, the attractor is classified as nonchaotic.
Furthermore, if the attractor has a fractal dimension, its dynamic is classified as strange nonchaotic.

\begin{figure}
    \centering
    \subfloat[\label{fig: mutual}]{%
        \includegraphics[width=0.24\textwidth]{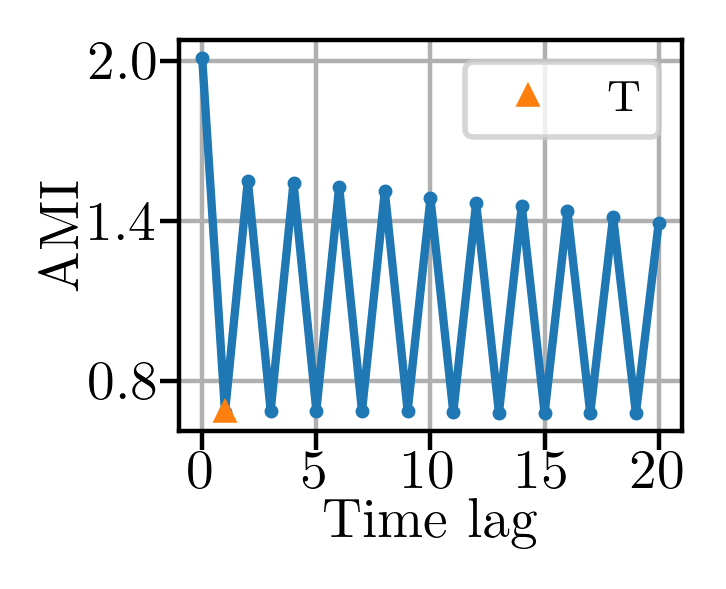}}
\centering   
    \subfloat[\label{fig: FNN}]{%
        \includegraphics[width=0.24\textwidth]{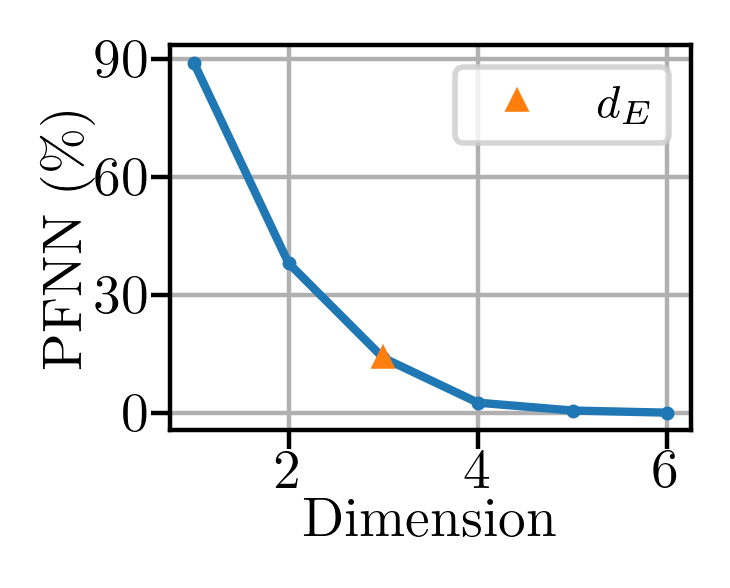}}
    \\\vspace{-1.5cm}
    \subfloat[\label{fig: 3d}]{%
    \includegraphics[width=0.48\textwidth]{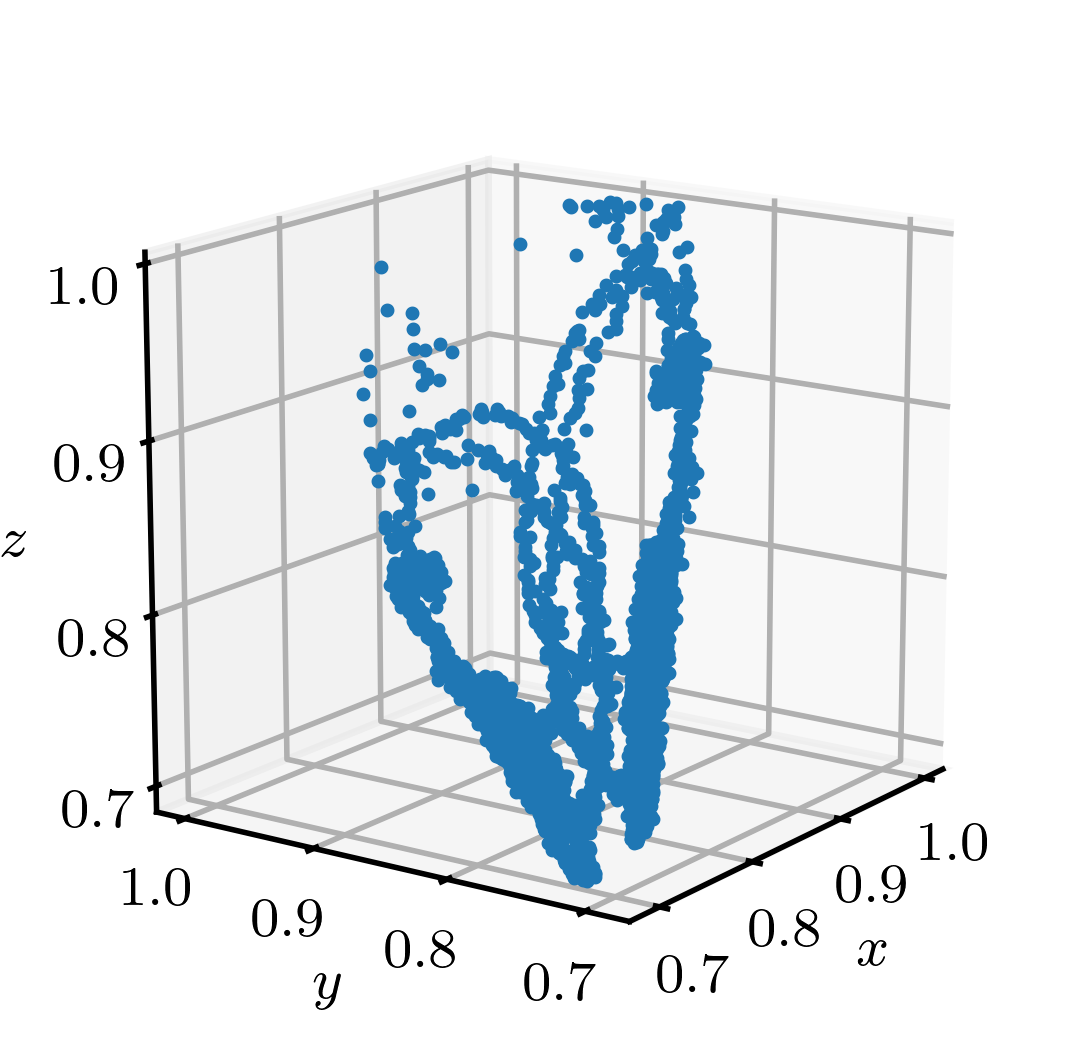}}

    \caption{Reconstruction of the attractor for a time series of difference between successive pulses. (a) Average mutual information (AMI) as a function of time lag. The blue dots mark the value of the average mutual information and the orange triangle the value of the first minimum, that in this case equals 1. (b) Proportion of false nearest neighbors. The blue dots are the value of the proportion of FNN at each dimension. The orange triangle marks the embedding dimension, $d_E$, that in this case is equal to 3. (c) Reconstructed attractor. The axis x, y and z are $s(n)$, $s(n+1)$ and $s(n+2)$ components of equation \eqref{eq:attractor} in arbitrary units. }
    \label{fig: attractor}
\end{figure}

We calculate the average mutual information, the proportion of FNN, and the Lyapunov exponents for each time series using the open source algorithms provided by Time Series Analysis (TISEAN) package \cite{tisean}.
For 10$\%$ of the total number of the analyzed time series, the mutual information could not be calculated, and $T$ was set to 1.
Figure \ref{fig: mutual} presents an example of the computed average mutual information for a time series of differences between successive pulses.
In this example, the first minimum of mutual information occurs at a time lag of 1, after which the mutual information shows oscillatory pattern.
This behavior reflects the typical pattern observed in all time series where mutual information can be computed.
The embedding dimension is usually defined as the dimension where the proportion of FNN first drops below 15$\%$ and continues to decrease.
A total of 32$\%$ of the time series meet this criterion and have a well-defined $d_E$. 
The remaining 68$\%$ of the time series, for which the dimension $d_E$ could not be determined, were excluded from further analysis in this study.
We attribute this to the challenges in calculating FNN due to the noise in the experimental data. 
Figure \ref{fig: FNN} shows an example of the proportion of FNN calculated as a function of dimension for the same example time series.
In this case, $d_E$ is equal to 3.
This behavior also represents the typical pattern of proportion of FNN in all time series where the embedding dimension is determinate.

Figure \ref{fig: 3d} shows the reconstructed attractor for the example time series, calculated using equation \eqref{eq:attractor}.
The figure clearly shows that the trajectories in the phase space are confined to a bounded region and do not expand to fill the entire space. 
This observation, which holds true for all time series where a finite embedding dimension was determined, indicates the presence of an attractor.

For the 1,096 time series whose attractors can be reconstructed, we calculate their respective Lyapunov spectra. All these series have a negative sum of Lyapunov exponents, and 94$\%$ have $\lambda_{max} > 0$, meaning that the series are chaotic. The remaining 6$\%$ of the time series exhibite a $\lambda_{max} < 0$, indicating that they are nonchaotic. For example, the Lyapunov spectrum computed for the reconstructed attractor shown in Figure \ref{fig: 3d} is: $\lambda_1=-0.03, \lambda_2=-0.31, \lambda_3=-1.05$.
To verify the robustness of our analysis regarding Lyapunov exponents, a parameter sweep of the time delay ($+$5) and embedding dimension ($+$2) revealed no change in the sign of the Lyapunov exponents and variations below 5$\%$, confirming the robustness of the reconstructed dynamics.

\begin{figure}
\includegraphics[width=0.48\textwidth]{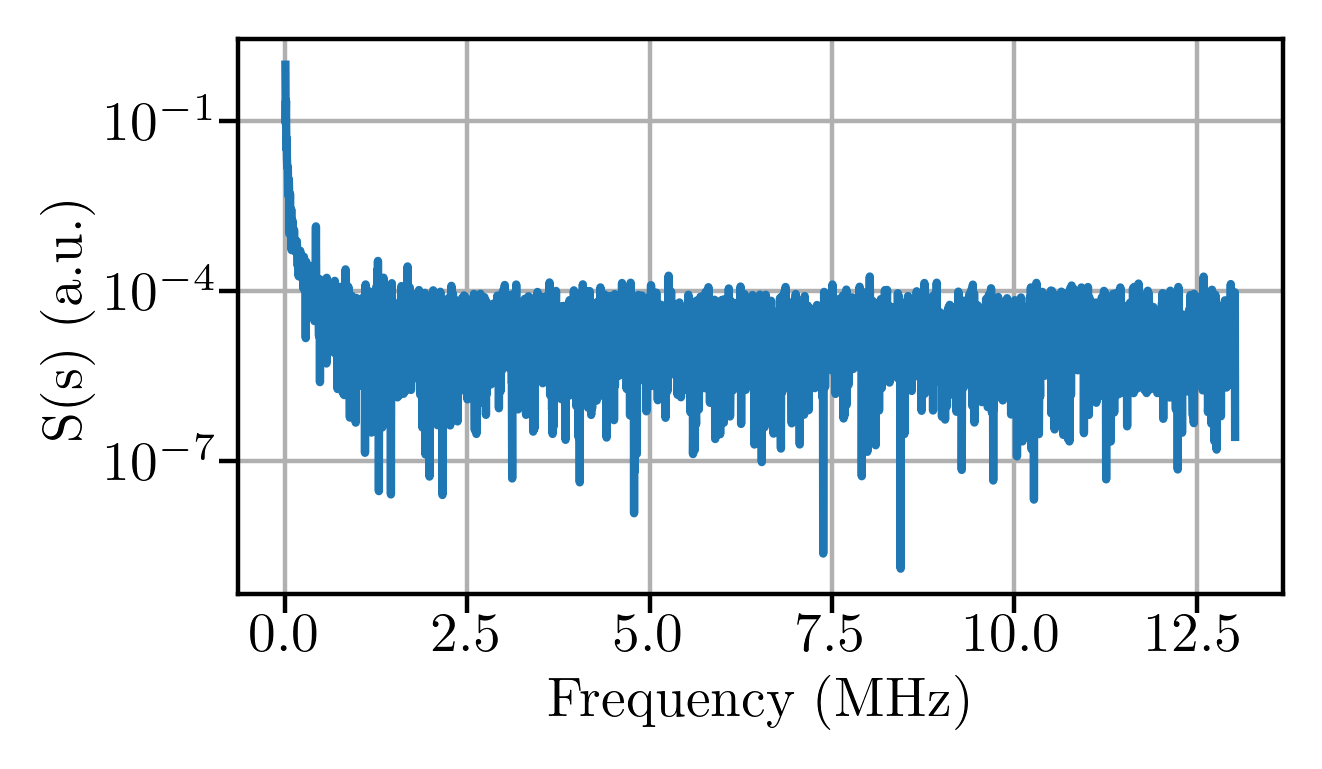}
\caption{\label{fig: fourier} Energy spectral density of a time series for a time series of difference between successive pulses.}
\end{figure}

We compute the energy spectral density for each nonchaotic time series.
The spectra of all these series exhibit a broad frequency range, indicating that the dynamics are not periodic and that the dynamical regimes are complex.
In Figure \ref{fig: fourier} we show the energy spectral density for a time series of the difference between successive pulses with $\lambda_{max} < 0$.

\begin{figure}
\includegraphics[width=0.48\textwidth]{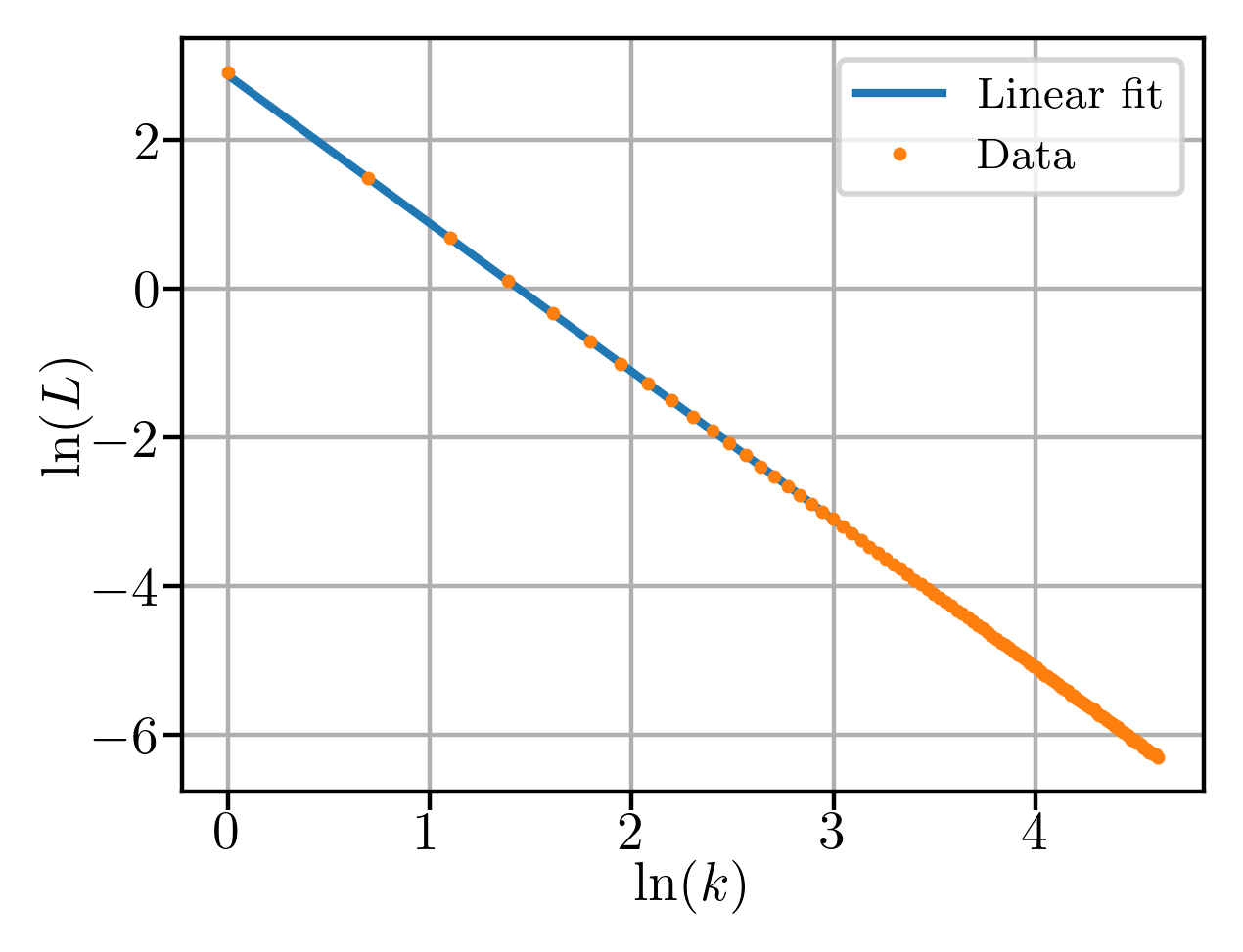}
\caption{\label{fig: fractal dim}         
        Logarithm of the curve length $L
        (k)$ calculated using Higuchi's method for a time series of the difference between successive pulses. The straight line represents the linear fit, and its slope is the negative of the estimated fractal dimension. In this case $d_f$ is 1.99.}
\end{figure}

We apply Higuchi's method to estimate the fractal dimension, $d_f$, for each time series with $\lambda_{max} < 0$ \cite{higuchi1988}.
This method constructs subsets of the time series using intervals of size $k$ and measures the length of each subset, $L(k)$.
If the relationship $L(k)\propto k^{-d_f}$ holds, the time series is fractal with dimension $d_f$.
To verify this relationship, we plot $\ln{L(k)}$ against $\ln{k}$ on a doubly logarithmic scale.
The data should align along a straight line with a slope $-d_f$. 
In our case, we fit the straight line to the data using the least squares procedure.
Higuchi's method recommends using a maximum value of $k$, $k_{max}$, that satisfies $k_{max} \leq N/10$, where $N$ is the length of the time series.
For all analyzed time series, we set $k_{max} = 100$, ensuring compliance with this condition.
Figure \ref{fig: fractal dim} displays a representative plot of this type, with an estimated fractal dimension of $d_f=1.99$. 
This example demonstrates a good fit, with a correlation coefficient $R =1.0$.
The fractal dimension can be also be estimated for 147 of the total chaotic time series, with values ranging from 1.90 to 2.00.
Similarly, among the nonchaotic time series, $d_f$ can be reliably determined for 8 cases, with values ranging from 1.86 to 1.99. 
In order to validate the Higuchi fractal dimension estimated from the eight experimental nonchaotic time series, a surrogate data analysis \cite{surro} was carried out.
For each series, 100 surrogate datasets were generated using both the random shuffle and the amplitude adjusted
Fourier transform (AAFT) methods.
In all cases, statistical testing revealed p-values below 0.0001, indicating that the original Higuchi dimensions differ significantly from those expected under the null hypotheses of independent identically distributed noise or purely linear stochastic processes.
These results provide strong evidence that the complexity quantified by the Higuchi method arises from intrinsic nonlinear dynamics rather than from trivial correlations or random fluctuations.

Therefore, we confirm that the underlying attractors of these 8 time series are strange nonchaotic, representing 0.2$\%$ of the total measured time series.
The time series for which the fractal dimension cannot be determined do not exhibit a well-defined $L(k)\propto k^{-d_f}$ relationship, likely due to the limitations of Higuchi's method when dealing with noisy data.
Consequently, their fractal dimension cannot be estimated, and we cannot confirm whether their structure are strange.
Nevertheless, it exhibits complex non periodic dynamics.
Table \ref{tab: summary attractors} presents a summary of the time series classifications.

\begin{table}[b!]
\caption{\label{tab: summary attractors}
Summary of the analyzed time series. The first column indicates the type of time series attractor. Where unclassified are time series for which $d_E$ could not be determined. The second column shows the number of time series, $M$, for each attractor type. The third column lists the range of time delay values observed, $T$, the fourth column displays the observed range of embedding dimension $d_E$, and the fifth column shows the range of $d_f$ values.}
\begin{ruledtabular}
\begin{tabular}{lcccc}
Attractor           & $M$    & $T$        & $d_E$             & $d_f$\\
\colrule
unclassified        & 2342  & $-$       & $-$               & $-$\\ 
chaotic             & 1029 & $1,...,17$   & $3,...,6$           & $1.90,...,2.00$\footnote{The fractal dimension was determined for 147 out of 1029 chaotic time series.}\\ 
nonchaotic         & 59   & $1$      & $3,...,5$           & $-$\\ 
strange nonchaotic & 8   & $1$       & $4,5$           & $1.86,...,1.99$\\
\end{tabular}
\end{ruledtabular}
\end{table}

To further characterize the reconstructed attractors with $\lambda_{max} < 0$, we calculated Kolmogorov complexity of each time series.
Kolmogorov complexity is a fundamental concept in algorithmic information theory that measures the computational resources needed to specify a binary sequence.
The Kolmogorov complexity, $K_c$, of a string $x$, or any object represented as a binary string, is the length of the shortest possible program (in a fixed programming language) that can produce $x$ when run on a universal Turing machine. 
Unfortunately, $K_c$ is incomputable, so we use an approximation to estimate its value.
$K_c$ is related to the incompressibility of the string and provides a measure of unpredictability and complexity in an information sense.
Each time series was binarized using its median value, and complexity was computed using the Mihailovic variant \cite{Dragu} of the Lempel-Ziv algorithm \cite{lempel_Ziv}.
$K_c$ is normalized to a scale where 0 corresponds to a perfectly predictable binarized time series (e.g., a periodic signal), and 1 represents a fully unpredictable (random) sequence.
Intermediate values indicate complex nonperiodic dynamics.
We find that the eight time series with all negative Lyapunov exponents and measured fractal dimensions exhibit $K_c$ values ranging from 0.1 to 0.4.
This intermediate range implies the time series is neither periodic nor random, but exhibits deterministic complexity which is a hallmark of SNAs. 

\section{Conclusions}
In summary, we present experimental evidence of the existence of a strange nonchaotic attractor in a passively Q-switched Nd:YVO$_4$ laser. 

We measure the laser output intensity for several positions of the Cr:YAG and, for each data set, generated time series of maximum pulse intensity and inter-pulse intervals, resulting in a total of 3,438 time series.
To analyze these time series, we reconstruct the underlying attractors by determining an optimal time delay and embedding dimension using mutual information and false nearest neighbors methods. 
We successfully reconstruct attractors for 32$\%$ of the time series.
For these attractors, we calculate the Lyapunov exponents and found that 30$\%$ of the total time series exhibited a negative sum of Lyapunov exponents and a positive maximum exponent, indicating chaotic behavior. 
In contrast, the remaining reconstructed attractors, representing 2$\%$ of the total time series, exhibited both a negative sum of Lyapunov exponents and a negative maximum exponent, confirming their nonchaotic nature.
We compute their fractal dimension using Higuchi’s method. We obtained a well-defined fractal dimension for 12$\%$ of the time series with $\lambda_{max} < 0$. 

The observed Kolmogorov complexity ($K_c$ = 0.1–0.4), combined with a fractal dimension between 1.86 and 1.99, further supports the identification of a strange non-chaotic attractor (SNA).
The intermediate $K_c$ values reflect deterministic complexity consistent with fractal geometry, while the negative Lyapunov exponents exclude chaotic dynamics.
This aligns with theoretical SNA signatures.
Therefore, we conclude that these time series exhibit strange nonchaotic behavior.

To the best of our knowledge, this represents the first experimental observation of a strange nonchaotic attractor in an optical system—specifically, a vanadate laser—under standard design and normal operating conditions.
The observation of a strange nonchaotic attractor in a Nd:YVO$_4$ laser is of interest not only from a fundamental perspective but also for potential applications.
SNAs combine a fractal geometry with the absence of exponential sensitivity to initial conditions, enabling complex yet robust dynamics.
Such properties can be exploited in precision metrology and optical sensing, where the system’s rich spectral response enhances sensitivity to weak perturbations while preserving stability.
In addition, the distinctive spectral characteristics of SNAs may be utilized in secure optical communications and signal processing, providing robustness against noise without the unpredictability of chaos.
Finally, the Nd:YVO$_4$ laser offers a controllable platform for exploring SNA dynamics, serving as an experimental model for similar phenomena in other physical systems.

\section*{Acknowledgments}
This work received support from the grants PUE 229-
2018-0100018CO and PIP 2022-00484CO CONICET (Argentina).

\bibliography{paper}

\end{document}